# Fractional Chern insulators in magic-angle twisted bilayer graphene

Yonglong Xie[1,2*‡], Andrew T. Pierce[1*], Jeong Min Park[2*], Daniel E. Parker[1], Eslam Khalaf[1], Patrick Ledwith[1], Yuan Cao[2], Seung Hwan Lee[1], Shaowen Chen[1], Patrick R. Forrester[1], Kenji Watanabe[3], Takashi Taniguchi[4], Ashvin Vishwanath[1], Pablo Jarillo-Herrero[2‡], Amir Yacoby[1‡]

[1]*Department of Physics, Harvard University, Cambridge, MA 02138, USA*
[2]*Department of Physics, Massachusetts Institute of Technology, Cambridge, MA 02139, USA*
[3]*Research Center for Functional Materials, National Institute for Material Science, 1-1 Namiki, Tsukuba 305-0044, Japan*
[4]*International Center for Materials Nanoarchitectonics, National Institute for Material Science, 1-1 Namiki, Tsukuba 305-0044, Japan*

*These authors contributed equally to this work.
‡Corresponding authors' emails: yxie1@g.harvard.edu, pjarillo@mit.edu, yacoby@g.harvard.edu

**Fractional Chern insulators (FCIs) are lattice analogues of fractional quantum Hall states that may provide a new avenue toward manipulating non-abelian excitations. Early theoretical studies[1–7] have predicted their existence in systems with energetically flat Chern bands and highlighted the critical role of a particular quantum band geometry. Thus far, however, FCI states have only been observed in Bernal-stacked bilayer graphene aligned with hexagonal boron nitride (BLG/hBN)[8], in which a very large magnetic field is responsible for the existence of the Chern bands, precluding the realization of FCIs at zero field and limiting its potential for applications. By contrast, magic angle twisted bilayer graphene (MATBG)[9–12] supports flat Chern bands at zero magnetic field[13–17], and therefore offers a promising route toward stabilizing zero-field FCIs. Here we report the observation of eight FCI states at low magnetic field in MATBG enabled by high-resolution local compressibility measurements. The first of these states emerge at 5 T, and their appearance is accompanied by the simultaneous disappearance of nearby topologically-trivial charge density wave states. Unlike the BLG/hBN platform, we demonstrate that the principal role of the weak magnetic field here is merely to redistribute the Berry curvature of the native Chern bands and thereby realize a quantum band geometry favorable for the emergence of FCIs. Our findings strongly suggest that FCIs may be realized at zero magnetic field and**

**pave the way for the exploration and manipulation of anyonic excitations in moiré systems with native flat Chern bands.**

The search for novel material systems exhibiting topological properties holds promise for the next generation of electronics. For example, band-structure engineering guided by theoretical predictions has enabled the realization of integer quantized Hall states at zero magnetic field[18–20], enabling new directions in spintronics and topological quantum computing. Likewise, extensive efforts have been directed toward engineering fractional Chern insulators (FCIs) —lattice analogues of fractional quantum Hall (FQH) states—in part because of their potential to manifest high-temperature topological order and to host non-abelian excitations at zero magnetic field. However, despite a large body of theoretical work[1–7], FCI states have proven exceptionally difficult to stabilize experimentally, as they not only require non-dispersive Chern bands, but also a particular quantum band geometry including a flat Berry curvature distribution. To date, FCI states have only been observed in Hofstadter bands of a bilayer graphene (BLG) heterostructure aligned with hexagonal boron nitride (hBN) at very large (~30 T) magnetic fields[8]. A key disadvantage of this platform is that its band topology fundamentally originates from the presence of the magnetic field, thus precluding the realization of FCIs in the zero-field limit.

On the other hand, moiré superlattices with native topological bands[13–17] provide a promising avenue to search for FCIs at zero magnetic field. In particular, the recent discovery of correlated Chern insulators in magic-angle twisted bilayer graphene (MATBG) down to zero magnetic field confirms the presence of intrinsic flat Chern bands[20–29] and thus raises the possibility of realizing FCIs in this system. Indeed, recent analytical considerations[30] and numerical calculations[31–33] have predicted FCI ground states in MATBG aligned with hBN. Importantly, these works also show the close competition between FCIs and other correlated phases such as charge density waves, and highlight the importance of Berry curvature distribution homogeneity and the quantum metric in stabilizing FCIs in MATBG. Here, we report the observation of eight FCI states at fractional fillings of the Chern bands in MATBG. The first of these states appear at 5 T in the range $3<\nu<4$, where the system is well-described by an isolated Chern band. We show that these FCI states appear as a result of the intrinsic band topology of MATBG and are stabilized by weak magnetic fields that create favorable quantum-geometric conditions for their emergence. The FCIs observed beyond this range, where the parent Chern states likely reacquire their multicomponent character, are more complex, likely due to the interplay between multiple degrees of freedom, and demonstrate the potential of MATBG for exploring novel emergent topological order.

To search for such topological ground states, we perform local electronic compressibility measurements on an MATBG device with a twist angle of ~ 1.06° (see Methods) using a scanning single electron transistor (SET). Our measurements of the inverse compressibility $d\mu/dn$ as a function of perpendicular magnetic field $B$ and moiré band filling factor $v$ reveal a large number of linearly-dispersing incompressible states (Fig. 1a and b) that can be classified by a pair of quantum numbers $(t, s)$ satisfying the Diophantine equation $v = t\phi/\phi_0 + s$, where $v$ is the filling factor at which the incompressible peak occurs, $\phi$ is the magnetic flux per moiré unit cell, and $\phi_0$ is the magnetic flux quantum. We observe in total five distinct classes of incompressible states. First, incompressible features with $t=0$ and integer $s \neq 0$ correspond to trivial correlated insulators (green line in Fig. 1b). Second, features with integer $t \neq 0$ and integer $s$ correspond to integer quantum hall states or Chern insulators (black lines in Fig. 1b), some of which have been identified as translation symmetry broken states resulting from unit cell doubling. The observed Chern insulators at zero magnetic field are the parent states essential for realizing more complex topological states in the zero-field limit. Finally, we observe three classes of gapped states with fractional $t$ and/or $s$, which we identify as charge density waves (CDWs—$t=0$ and fractional $s$), symmetry broken Chern insulators (SBCIs—integer $t \neq 0$ and fractional $s$) and FCIs (fractional $t$ and fractional $s$).

To demonstrate that our system provides the topological bands and strong correlations essential for the realization of FCIs, we focus on the range of filling factors near $v=3$, as in this density range the band structure can be best approximated by isolated Chern bands. Fig. 2a shows a measurement of inverse compressibility as a function of magnetic field for $2.5<v<4$ for $B<3$ T. In addition to the insulators emanating from $v=3$, we discover three new incompressible states that are stable down to zero magnetic field: the two non-dispersive states (0, 7/2) and (0, 11/3), which we classify as trivial CDWs, and the SBCI state (1, 8/3) (Fig. 2b). The fractional values of $s$ associated with these states strongly suggest electron-electron interactions spontaneously break the translation symmetry of the underlying moiré superlattice, likely resulting in new real-space charge order with an enlarged unit cell. In fact, a previous study[29] has shown that the appearance of a portion of the Chern insulators is likely a consequence of translation symmetry breaking via doubling of the unit cell. In this scenario, the Hartree potential favors filling states near the center of the mini-Brillouin zone, which in MATBG is also the region where the Berry curvature is highly concentrated (yellow trace in Fig. 2c). As a result, the system may favor forming one band that retains the original Berry curvature and therefore has $C=\pm 1$, along with a new $C=0$ band (Fig. 2d). Under this assumption, filling three of the four $C=0$ bands generated by unit cell doubling yields the (0, 7/2) state (Fig. 2f). Similarly, tripling the unit cell allows one $C=\pm 1$ band to give rise to a $C=\pm 1$ band accompanied by two $C=0$ bands (Fig. 2e). Sequentially filling the 12 reconstructed bands produces both the (1, 8/3) and (0, 11/3) states (Fig. 2f and g). Together, the observation of CDW and SBCI states at zero field establishes the presence of both intrinsic band topology and strong electron-electron interactions, and highlights the critical role of the non-uniform Berry curvature in stabilizing these two classes of states.

Remarkably, upon increasing the magnetic field to 5 T, we observe a different family of robust incompressible states that are parametrized by fractional values of both $t$ and $s$ (Fig. 3a and b), characteristic of FCIs. These states, (2/3, 10/3) and (1/3, 11/3), persist up to at least 11 T, and can be interpreted as lattice analogues of $v_c =1/3$ and $2/3$ FQH states from the final $C=-1$ band populated upon electron-doping the (1, 3) Chern insulator, where $v_c$ is the filling factor of the partially-filled Chern band (Fig. 3c). Because these states do not require breaking of the translation symmetry of the moiré superlattice, despite their fractional $s$, they are referred to below as symmetry-preserving FCIs. These two states are expected to exhibit fractional quantized Hall conductance according to the Streda formula and hence support quasiparticle excitations with fractional charge e/3[34]. Integrating the inverse compressibility d$\mu$/d$n$ with respect to the electron density allows us to directly extract the steps in chemical potential $\Delta\mu$ associated with each of the observed CDW and FCI states (Fig. 3d). Because the chemical potential is defined with respect to electrons, $\Delta\mu$ must be multiplied by the ratio of the quasiparticle charge to the electron charge, yielding energy gaps of about 50±20 μeV (~0.6 K) for both FCI states, roughly in agreement with the estimate of 0.01$U$ from a recent exact diagonalization study[33], where $U$ is the strength of Coulomb interaction. The same study also argues that, because the spin polarization of the valley-polarized Chern band is unknown, the FCI states can be either isospin-polarized Laughlin states or multicomponent states depending on the detailed quantum-geometric properties of the system. While our measurements are not capable of directly distinguishing between single and multicomponent ground states, we note that the gaps associated with both FCIs are much smaller than the spin Zeeman energy scale $E_Z=g\mu_B B$ (assuming $g=2$) and depend very weakly on $B$, suggesting that the charged excitations of both states likely do not require a spin flip. The sudden appearance of the FCIs and disappearance of the CDWs indicates close competition between these two phases, with the magnetic field driving the transition yet leaving the band topology unaltered.

To understand the transition from a CDW-dominated to an FCI-dominated regime, we begin by observing that these two classes of ground states place very different constraints on the quantum-geometric properties of the underlying band structure of MATBG. For the CDW ground states to emerge, the Berry curvature of the flat bands must be strongly concentrated near the center of the mini-Brillouin zone to take advantage of the Hartree potential[29]. However, bands with sufficiently nonuniform Berry curvature are known to disfavor FCI ground states[30–33]. The observed transition therefore suggests that the applied magnetic field in the experiment serves primarily to reduce the intrinsic Berry curvature inhomogeneity within the partially-filled Chern band, unlike in the hBN/BLG system where the applied field is needed to produce Chern bands in the first place. To estimate the amount of Berry curvature inhomogeneity the FCI ground states can tolerate, we note that exact diagonalization studies[31–33] indicate the presence of a transition between CDW and FCI ground states as a function of $w_0/w_1$, where $w_0$ is the interlayer tunneling matrix element at the AA-stacked regions and $w_1$ is that at the AB-stacked regions. This ratio is known to strongly alter the Berry curvature distribution within the flat bands of MATBG. According to these works, the transition occurs near $w_0/w_1 \approx 0.7$, as has been confirmed by a recent DMRG study[35]. The ground-state dependence on $w_0/w_1$ therefore gives a means of parametrizing the dependence of the FCI ground states on the Berry curvature inhomogeneity, which we characterize using the quantity $\sigma(F)$, the mean standard deviation of the Berry curvature over the mini-Brillouin zone. These results allow us to estimate an upper bound on the allowable Berry curvature inhomogeneity $\sigma_c(F)$ to be in the range of 1.4 to 2.2, depending on the model parameters, below which the quantum geometry of the system is favorable for the emergence of FCIs (Fig. 3e). For simplicity, we choose $\sigma_c(F) = 1.8$ for the discussion below. For realistic MATBG samples, $w_0/w_1$ is estimated to be around $0.8$[25,36,37], yielding large values of $\sigma(F) \sim 3$, consistent with our observation of CDW states at zero magnetic field. Thus, the absence of FCI ground states at zero magnetic field in our device can be understood to result from the large values of $\sigma(F)$ present in MATBG.

Having established the critical role of $\sigma(F)$ in determining the many-body ground state, we now examine its evolution as a function of magnetic field by analyzing the Hofstadter spectrum of the continuum model of MATBG aligned with hBN (see Methods). We find that increasing the magnetic field reduces $\sigma(F)$ monotonically (Fig. 3f), with $\sigma(F)$ vanishing as $\phi/\phi_0 \rightarrow 1$. To estimate the value of magnetic field at which the FCI ground state becomes favorable, we identify the magnetic field, $B_c$, at which $\sigma(F)$ falls below the critical value $\sigma_c(F) \approx 1.8$. For realistic values of $w_0/w_1 \sim 0.8$, our calculations find that $\sigma(F)$ is reduced below $\sigma_c$ starting at $\phi/\phi_0 \sim 1/5$ or $B_c \sim 5.4$ T, in good agreement with the magnetic field at which the (2/3, 10/3) and (1/3, 11/3) are observed to appear experimentally. We emphasize that due to the sharp decrease of $\sigma(F)$ with field, the critical field $B_c$ is not sensitive to the precise choice of $\sigma_c(F)$. Combining the bound $\sigma_c(F)$ estimated from many-body ground state analyses[31–33,35] at zero magnetic field with our calculations of $\sigma(F)$ as a function of $B$ and $w_0/w_1$ allows us to sketch a phase diagram at $v_c = 1/3$ (Fig. 3g). Our calculations also demonstrate that the FCI is adiabatically connected to the FQH state at $\phi/\phi_0 = 1$, where the band geometry reduces to that of the lowest Landau level. However, unlike the case of the usual FQH states or of FCIs occurring within partially-filled Hofstadter bands in a BLG/hBN heterostructure, in which the Berry curvature is supplied by the Landau levels or Chern bands that form in a magnetic field, the FCIs observed here fundamentally stem from zero-field Chern insulator parent states, and the only role of the magnetic field is to flatten the Berry curvature. Therefore, only a weak magnetic field of less than 20% of a magnetic flux quantum per moiré unit cell is required to stabilize the FCIs by reducing $\sigma(F)$ below $\sigma_c(F)$.

Outside the density range 3<v<4 the system recovers additional degrees of freedom and thus permits more possible competing ground states at fractional fillings. In particular, we observe six additional FCIs—along with numerous SBCIs with denominators of s as large as 10 (Extended Data Fig. 1)—at other fillings at slightly higher values of magnetic field, particularly on the hole side (Fig. 4), most of which show values of $\Delta\mu$ comparable to those of their counterparts near v=3 (Extended Data Fig. 2). We emphasize that our measurements unambiguously identify these states as FCIs purely based on the Streda formula independent of their exact nature and origin, on which we speculate below. As in the case of the FCIs observed for $3 < v < 4$, several of these additional states likely correspond to symmetry-preserving FCIs (Fig. 4a-b). For example, we interpret the state (-4/3, -5/3) as arising from a $v_c = 1/3$ FCI formed within the C = -1 band populated upon electron-doping the (-1, -2) Chern insulator, similar to the (2/3, 10/3) and (1/3, 11/3) states described above. In addition, unlike the aforementioned states, the observed (-8/5,11/10) and (-7/3, 2/9) states (Fig. 4c and d) have $s$ with denominator a multiple of that of $t$, rather than being equal, suggesting that each unit cell only binds a fraction of an electron charge and thereby corresponding to a symmetry-broken FCI. Specifically, the (-8/5, 11/10) can result from doping the (-2,1) Chern insulator with a $v_c = 2/5$ FCI that quadruples the unit cell and thus only contributes to a change in $s$ of 1/10. This interpretation is further supported by the fact that the (-2, 1) Chern insulator is a state that breaks translation symmetry, and thus might naturally be expected to also support symmetry-broken FCIs. We note that such symmetry-broken FCI states have no analogue in the FQH system. Most intriguingly, we also find evidence of FCI states with coprime denominators of $t$ and $s$ that as a result cannot be described as either symmetry-preserving or symmetry-breaking FCIs. The emergence of these exotic many-body ground states may originate from complex interplay between spin, valley and spatial symmetry.

The observation of FCIs in MATBG reported here leaves open many theoretical and experimental questions. An interesting and straightforward direction is to identify the quasiparticle charge associated with these FCI states, especially those that have no analogues in the FQH system. The competition between FCIs and nearby CDWs and SBCIs may provide a new setting for the study of quantum phase transitions. Importantly, our work establishes the applied magnetic field as a novel tuning knob for the Berry curvature distribution, and indicates close proximity to zero-field FCIs in the flat bands of MATBG. Thus, a pressing experimental task is to develop means of reducing $w_0/w_1$ in MATBG and to explore alternative platforms beyond MATBG that suffer less from Berry curvature inhomogeneity, which would enable the realization of FCIs at zero magnetic field and offer new opportunities for the creation of next-generation topological quantum devices.

## Methods

**Sample preparation.** The MATBG device used in this study was fabricated using the "tear-and-stack" technique described in Ref.[38,39], and is the same as the one in Ref.[29]. Briefly, the monolayer graphene and hBN flakes were first exfoliated on $SiO_2$/Si substrates and subsequently screened with optical microscopy and atomic force microscopy. We use a PC/PDMS stamp on a glass slide to sequentially pick up the flakes. The resulting stack is released on the pre-stacked hBN-on-Pd/Au back gate. The device geometry was defined by electron-beam lithography and reactive ion etching. Cr/Au electrical contacts to MATBG were made by the standard edge-contact method.

**Compressibility Measurements.** All compressibility measurements were made in a $^3$He cryostat. The SET tips were fabricated using a procedure described elsewhere[40]. Compressibility measurements were performed using DC and AC protocols similar to those described in Ref[29,40]. Compared to Ref[29], a longer integration and voltage-ramping time and a slightly smaller tip-sample distance were used to further improve the signal to noise.

**Procedure of determining the slope and the intercept for the fractional states.** We determine the quantum numbers ($t$, $s$) of the incompressible states by first identifying the peaks associated with each state and performing a linear fit to obtain their slope and intercept. To more accurately confirm the fractional values of $t$ and $s$ and mitigate the error due to effects of quantum capacitance[8], we use the fitted slope and intercept of the nearby Chern and correlated insulators to obtain local estimates of $t$ and $s$. Based on the converted values of $t$ and $s$, we assign the corresponding fractions for $t$ and $s$ by identifying those with the smallest denominator possible (up to 10) within the 95% confidence interval and favor the fractions of $t$ and $s$ that share the same denominators (Extended Data Fig. 3).

**Hofstadter spectrum.** We model the system using the Bistritzer-MacDonald model[9] with a twist angle of $\theta = 1.06°$, and account for the gap at charge neutrality observed in the experiment by including a sublattice splitting of 30 meV. The interlayer tunneling parameter $w_1$ is set to 110 meV. In the range $3<\nu<4$, the system can be approximated by a single Chern band, and we thus consider a single fermion species in our calculation. We obtain the Hofstadter spectrum following Ref.[41–43], which is shown in Extended Data Fig. 4. As dictated by the Streda formula, the top $C=-1$ band—the parent state of the FCI—is separated by a gap. Complete details of the model are given in the Supplementary Information.

**Quantum geometry**. The stability of FCI is closely related to the quantum geometry of the MATBG band structure. One key figure of merit is the standard deviation of the Berry curvature distribution over the Brillouin zone. In the presence of a magnetic flux $\phi = \frac{p}{q}\phi_0$, there are $2q$ bands in the Hofstadter spectrum, where $q - Cp$ bands are filled. Here, we present a natural multi-band generalization for the standard deviation of the Berry curvature, which is continuous, gauge invariant and reduces to the expected values at $\frac{\phi}{\phi_0} \to 0$. At a given magnetic flux $\phi$, let $\mathcal{P}_k = \sum_{a=1}^{N}|u_k><u_k|$ be the projector to the top $C=-1$ band. The $U(N)$ non-Abelian Berry curvature is defined as $F_{ab}(k) = -2NA\text{Im}(<\partial^x u_{ka}|1 - \mathcal{P}_k|\partial^y u_{kb}>)$, where $\partial^\mu = \frac{\partial}{\partial k^\mu}$ and $A = A_0\phi/\phi_0$ is the area of the magnetic Brillouin zone. We note that the non-standard normalization $NA$ is necessary for gauge-invariant quantities to be continuous functions of magnetic field. A semi-analytic formula for $F$, which is both numerically stable and accounts for the intrinsic geometry, is derived in the Supplementary Information. To evaluate expectation values of the Berry curvature distribution, we define the trace operator $Tr[O] = (NA)^{-1}\sum_{b=1}^{N}\int d^2k\, O_{bb}(k)$ so that $Tr[Id]=1$, where $Id$ is the identity operator. The Chern number $C = Tr[F/2\pi]$ is the mean of the distribution, up to $2\pi$. The standard deviation of the Berry curvature is then defined as $\sigma(F) = \sqrt{Tr[(\frac{F}{2\pi} - C)^2]}$.

## Data availability

The data that supports the findings of this study are available from the corresponding authors upon reasonable request.

## Code availability

The codes that supports the findings of this study are available from the corresponding authors upon reasonable request.

## Acknowledgements

We acknowledge discussions with Ady Stern and Yarden Sheffer. This work was primarily supported by the U.S. Department of Energy, Basic Energy Sciences Office, Division of Materials Sciences and Engineering under award DE-SC0001819. Fabrication of samples was supported by the U.S. Department of Energy, Basic Energy Sciences Office, Division of Materials Sciences and Engineering under award DE-SC0019300 and ARO Grant No. W911NF-14-1-0247. Help with transport measurements and data analysis were supported by the National Science Foundation (DMR-1809802), and the STC Center for Integrated Quantum Materials (NSF Grant No. DMR-1231319) (Y.C.). P.J-H acknowledges support from the Gordon and Betty Moore Foundation's EPiQS Initiative through Grant GBMF9643. A.T.P. acknowledges support from the Department of Defense through the National Defense Science and Engineering Graduate Fellowship (NDSEG) Program. Y.X. and S.C. acknowledge partial support from the Harvard Quantum Initiative in Science and Engineering. A.T.P., Y.X and A.Y. acknowledge support from the Harvard Quantum Initiative Seed Fund. AV was supported by a Simons Investigator award and by the Simons Collaboration on Ultra-Quantum Matter, which is a grant from the Simons Foundation (651440, AV). EK was supported by a Simons Investigator Fellowship, by NSF-DMR 1411343, and by the German National Academy of Sciences Leopoldina through grant LPDS 2018-02 Leopoldina fellowship. P.R.F. acknowledges support from the National Science Foundation Graduate Research Fellowship under Grant No. DGE 1745303. This research is funded in part by the Gordon and Betty Moore Foundation's EPiQS Initiative, Grant GBMF8683 to D.E.P. K.W. and T.T. acknowledge support from the Elemental Strategy Initiative conducted by the MEXT, Japan, Grant Number JPMXP0112101001, and JSPS KAKENHI Grant Number JP20H00354. This work was performed, in part, at the Center for Nanoscale Systems (CNS), a member of the National Nanotechnology Infrastructure Network, which is supported by the NSF under award no. ECS-0335765. CNS is part of Harvard University.

## Author Contributions

Y.X., A.T.P., J.M.P., P.J.-H. and A.Y. conceived and designed the experiment. A.T.P. and Y.X. performed the scanning SET experiment and analyzed the data with input from A.Y. J.M.P., Y.C. and P.J.-H. designed and provided the samples and contributed to the analysis of the results. D.P., E.K., P.L. and A.V. performed the theoretical analysis. K.W. and T.T. provided hBN crystals. All authors participated in discussions and in writing of the manuscript.

## Competing interests

The authors declare no competing interests.

# Figure 1

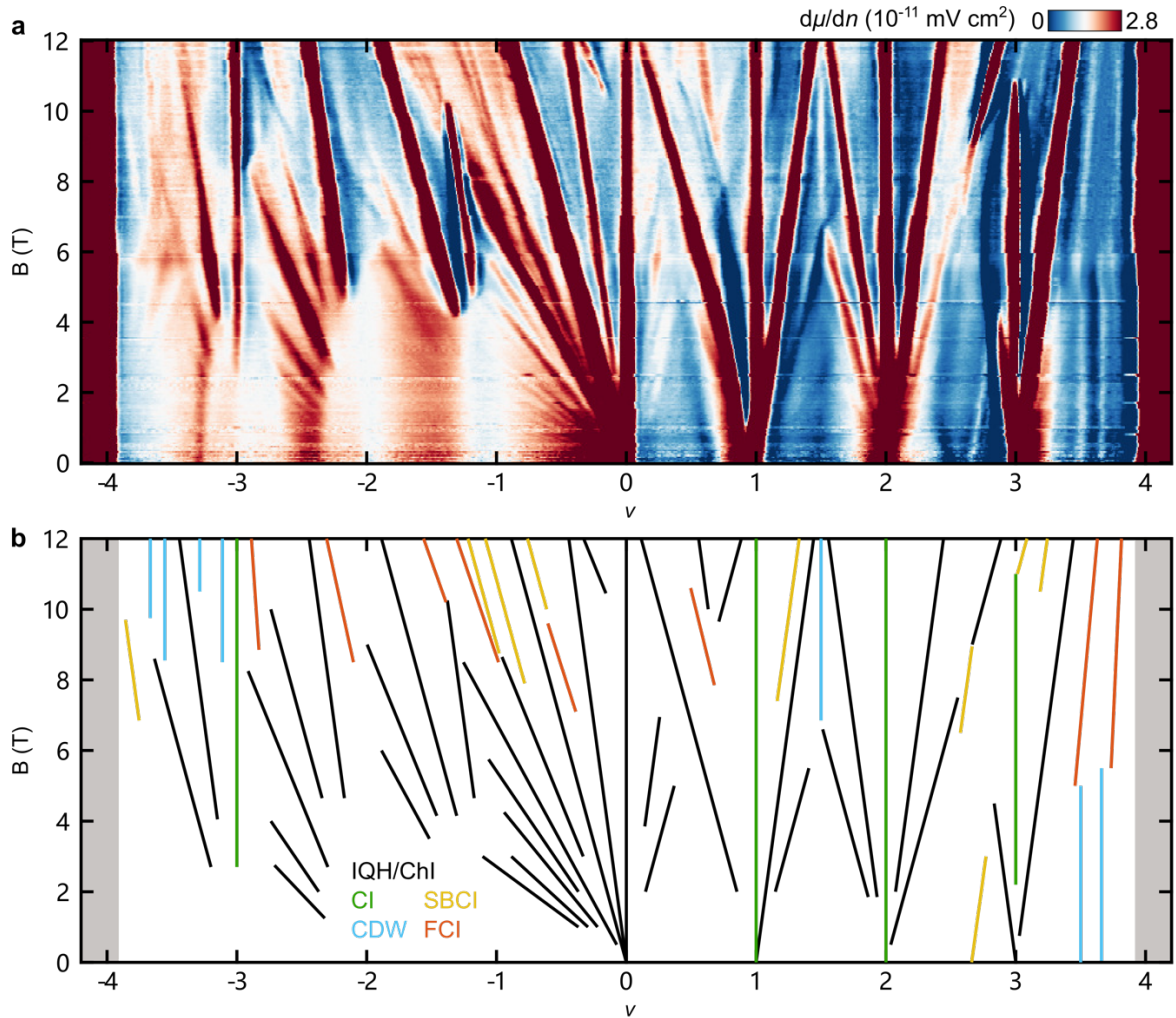

**Fig. 1 | Incompressible states with fractional quantum numbers in MATBG. a,** Local inverse compressibility d$\mu$/d$n$ measured as a function of magnetic field $B$ and electrons per moiré unit cell $v$. **b,** Wannier diagram identifying the incompressible peaks present in panel a. Black lines correspond to Chern insulators (ChIs) and integer quantum Hall (IQH) states; green lines correspond to correlated insulators (CIs) emanating with nonzero integer $s$ and $t$=0; blue lines correspond to charge density waves (CDWs) with integer $t$=0 and fractional $s$; yellow lines correspond to symmetry-broken Chern insulators (SBCIs) with nonzero integer $t$ and fractional $s$; and orange lines correspond to fractional Chern insulators (FCIs) with fractional $t$ and fractional $s$. Grey shaded regions correspond to the gaps to the remote bands.

# Figure 2

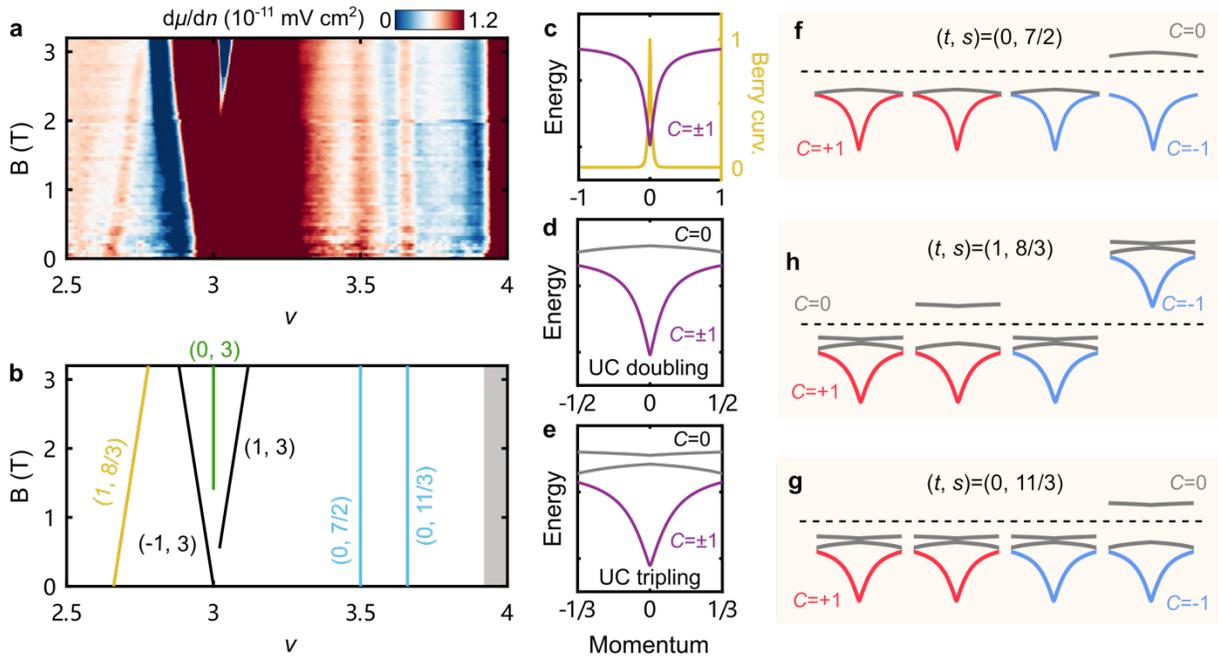

**Fig. 2 | Density wave states at low magnetic field for 2.5<ν<4. a,** Local inverse compressibility d$\mu$/d$n$ between ν=2.5 and 4 for $B$=0 T to 3 T. **b,** Schematic Wannier diagram corresponding to the states observed in **a** colored according to the classification used in Fig. 1b. **c,** Band energy (purple) and Berry curvature (yellow) along a path through the $\Gamma$ point in the first mini-Brillouin zone. Zero momentum corresponds to the $\Gamma$ point. **d,** Schematic band structure in the case of unit cell doubling resulting in a $C=\pm1$ band accompanied by a new $C=0$ band. **e,** Schematic band structure in the case of unit cell tripling resulting in a $C=\pm1$ band accompanied by two new $C=0$ bands. **f-g,** Schematic depiction of band fillings in the case of unit cell doubling **(f)** and unit cell tripling **(h-g)** needed to produce the density wave states observed in **a**.

# Figure 3

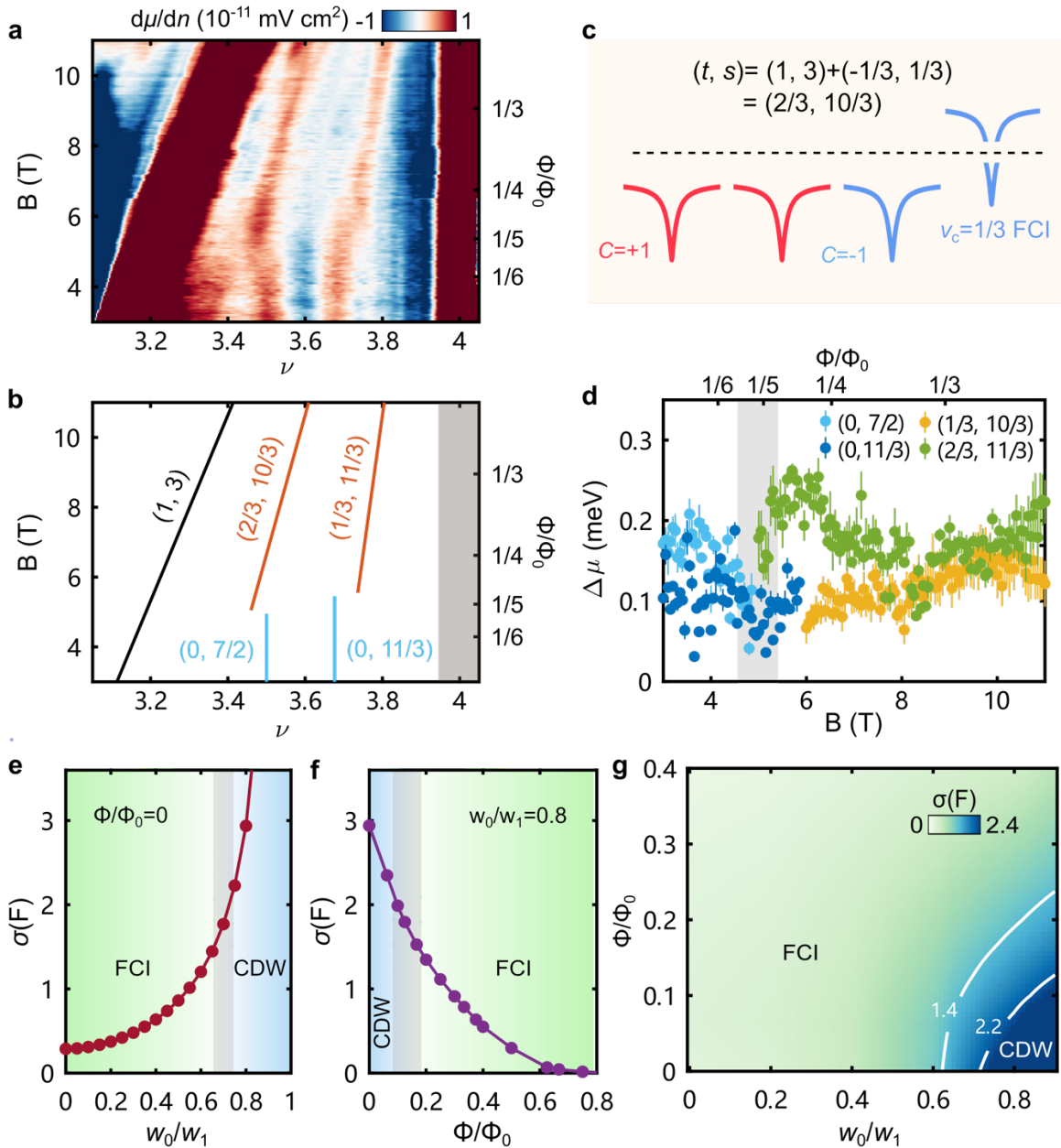

**Fig. 3 | Fractional Chern insulators in a weak magnetic field. a,** Local inverse compressibility d$\mu$/d$n$ between $\nu$=3 and 4 for $B$=3 T to 11 T. **b,** Schematic Wannier diagram corresponding to the states observed in **a** colored according to the classification used in Fig. 1b. Light blue and orange lines denote the CDWs and FCIs, respectively. The grey shaded region marks the energy gap to the remote band. **c,** Schematic depiction of band fillings that leads to the $(t, s)$=(2/3, 10/3)

FCI observed in **a**, which corresponds to a $\nu_c = 1/3$ FCI state from the final $C=-1$ band populated upon electron-doping the (1, 3) Chern insulator. **d,** Chemical potential steps $\Delta\mu$ associated with the CDW and FCI states observed in **a** obtained by integrating the inverse compressibility $d\mu/dn$. **e,** Calculated average Berry curvature deviation $\sigma(F)$ from the continuum model as a function of $w_0/w_1$. The grey shaded region corresponds to $w_0/w_1 = 0.65$ to 0.75, the range in which the transition from FCI to CDW occurs. This $w_0/w_1$ range allows us to estimate the range of values $\sigma_c(F) = 1.4$ to 2.2 below which the FCI is favorable. **f,** Calculated average Berry curvature deviation $\sigma(F)$ as a function of magnetic field for $w_0/w_1 = 0.8$. **g,** Schematic phase diagram constructed as a function of $w_0/w_1$ and magnetic field in units of $\phi_0$. The white lines indicate the contours $\sigma_c(F) = 1.4$ and $\sigma_c(F) = 2.2$ that define the region where the phase boundary between the FCI and CDW ground states is expected.

# Figure 4

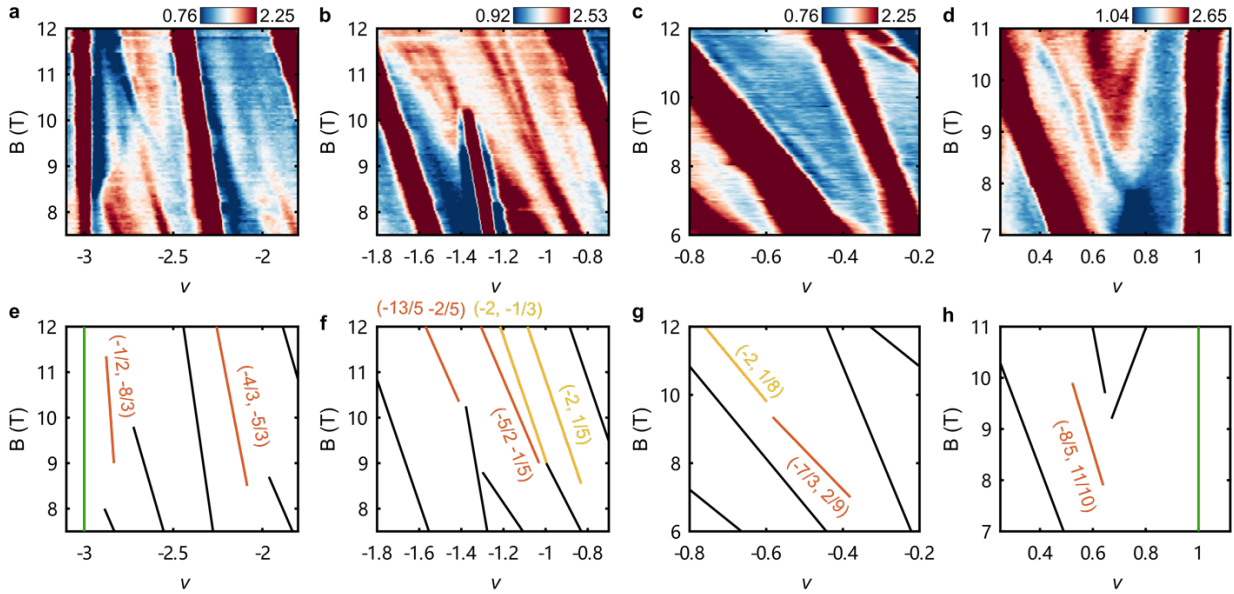

**Fig. 4 | Additional fractional Chern insulators at higher magnetic field. a-d,** Measurements of d$\mu$/d$n$ in various density ranges between 6 and 12 T showing additional FCIs and SBCIs. **e-g,** Schematic Wannier diagrams corresponding to the states observed in **a-d** colored according to the classification used in Fig. 1b. Green, yellow and orange lines denote the CIs, SBCIs and FCIs, respectively.

# Extended Data Figure 1

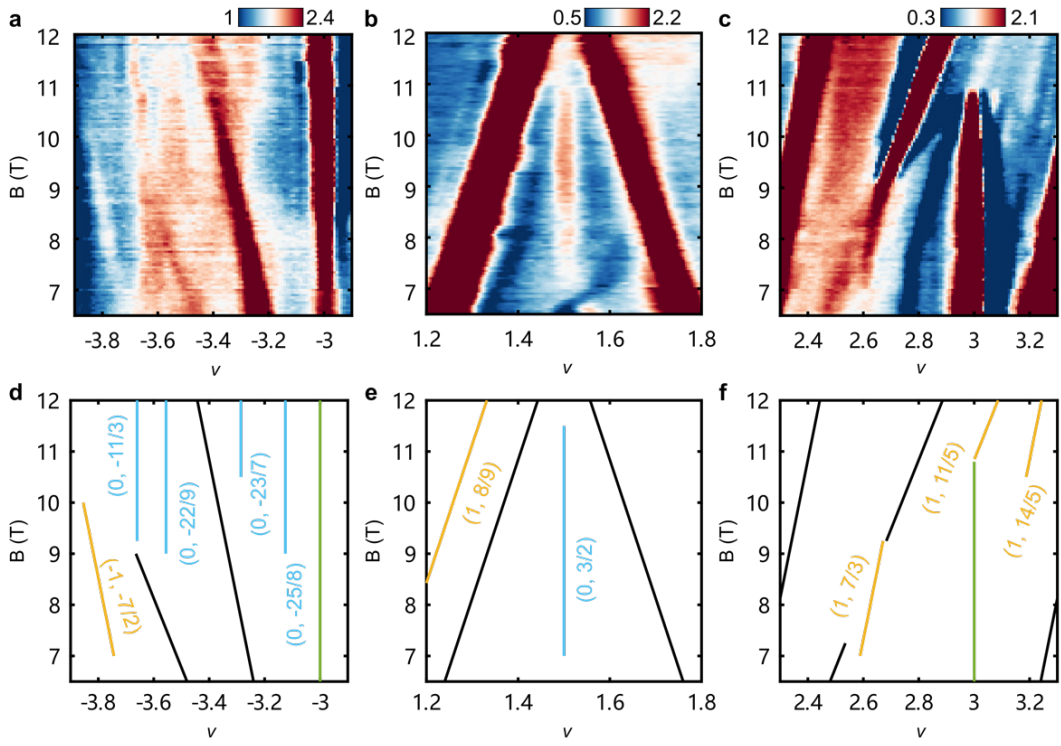

**Extended Data Fig. 1 | Additional CDW and SBCI states at higher magnetic field. a-d,** Measurements of d$\mu$/d$n$ in various density ranges between 6.5 and 12 T showing additional CDWs and SBCIs. **e-g,** Schematic Wannier diagrams corresponding to the states observed in **a-d** colored according to the classification used in Fig. 1b. Light blue and yellow lines denote the CDWs and SBCIs, respectively.

# Extended Data Figure 2

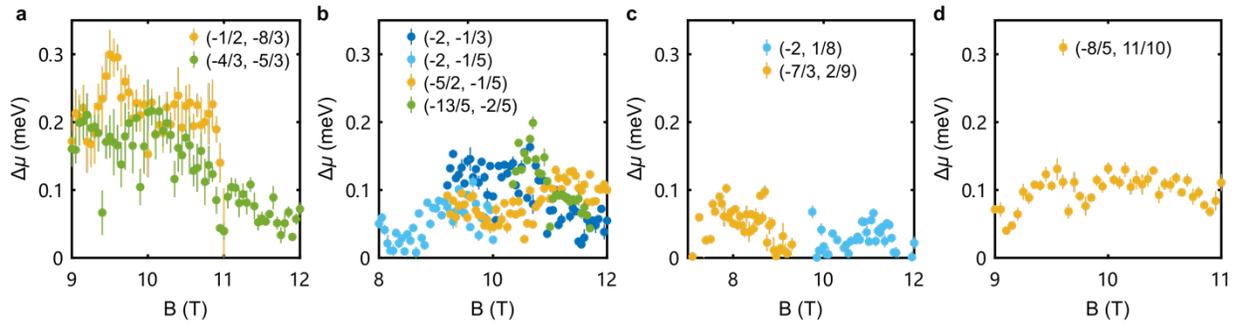

**Extended Data Fig. 2 | Energy gaps of additional FCI and SBCI states. a-d,** Chemical potential steps $\Delta\mu$ of the FCI (yellow and green circles) and SBCI (light and dark blue circles) states shown in Fig. 4.

# Extended Data Figure 3

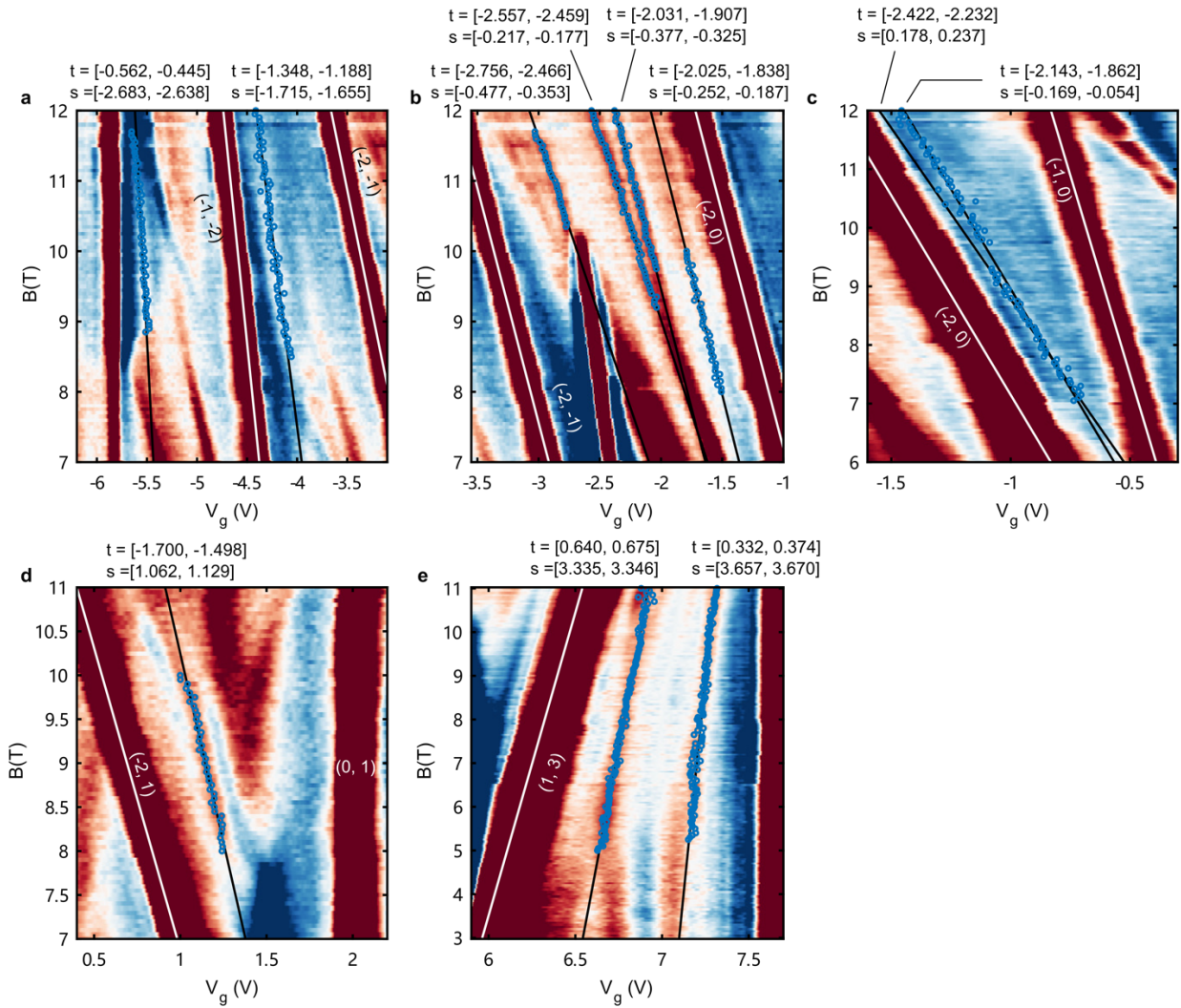

**Extended Data Fig. 3 | Fits to (*t*, *s*) for FCIs and SBCIs. a-e,** Incompressible peak locations (blue circles) associated with FCI and SBCI states. Black lines mark the results of linear fits. The fitted slope of nearby Chern insulators were used to convert the parameters to (*t*, *s*), the values of which are shown in the brackets with 95% confidence intervals.

# Extended Data Figure 4

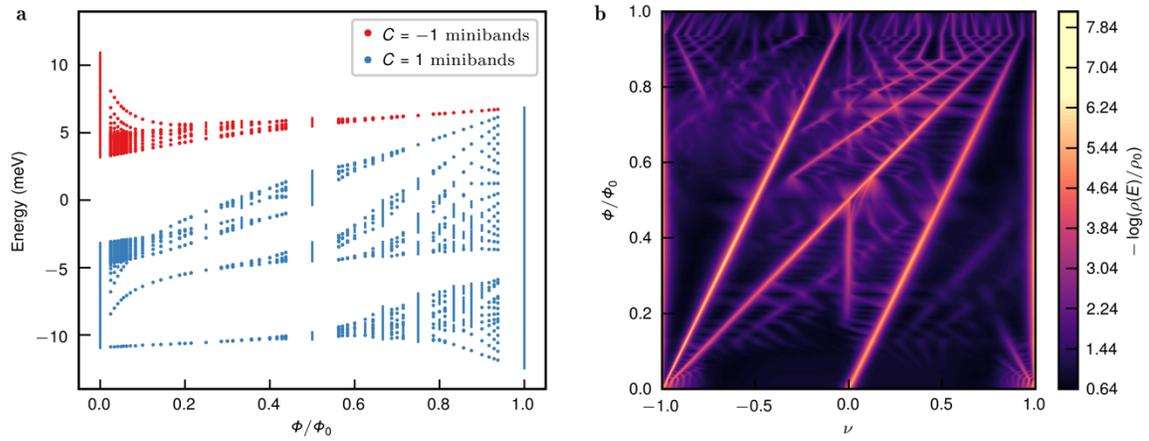

**Extended Data Fig. 4 | Hofstadter spectrum, a,** Calculated spectrum of the narrow bands of MATBG at finite magnetic field. The bands are colored according to their Chern number at zero magnetic field. **b,** Wannier plot corresponding to the spectrum in **a**.